\documentclass[epsf]{aa}
\usepackage{graphics}
\begin{document}

   \thesaurus{11.04.1; 11.05.2; 11.06.1; 12.03.3}

   \title{Imaging around PC1643+4631A at the Lyman limit }
   \author{I. Ferreras\inst{1,2}, 
	   N. Ben\'\i tez\inst{1,2} \and
	   E. Mart\'\i nez-Gonz\'alez\inst{1}}
   \offprints{I. Ferreras}
   \institute{Instituto de F\'\i sica de Cantabria, 
		CSIC--Universidad de Cantabria. 
		Fac. de Ciencias, Av. los Castros s/n, 39005 Santander, Spain.
                E-mail: ferreras@ifca.unican.es
	\and
	      Departamento de F\'\i sica Moderna. Universidad de
		Cantabria.
   }
   \date{Received ; accepted }

   \titlerunning{Imaging around PC1643+4631A}
   \authorrunning{I. Ferreras et al.}
   \maketitle
   \begin{abstract}
	We present a list of candidates for high redshift late--type
	galaxies in the field around the $z=3.79$ quasar PC1643+4631A.
	Deep $U$,$V$ and $R$ imaging has been used to search for objects
	with a strong Lyman break between $U$ and $V$, characteristic
	of galaxies with high hydrogen column densities at $z\sim 3$.
	A further study of the red objects detected by Hu \& Ridgway
	(\cite{hu}) has been done, allowing the temptative identification of 
	many of them as low redshift ($z\sim 0.4$) elliptical galaxies.
   \end{abstract}

\section{Introduction}

One of the most efficient methods of selecting high redshift galaxy 
candidates involves the detection of Lyman break systems. This break
comes from the likelihood of absorption due to intervening H I along the line 
of sight and --- for many galaxies --- also from the instrinsic continuum break
in the spectrum of the galaxy (Bruzual \& Charlot \cite{bruz}). The search 
is done with two broad band filters that fork this break and a third
one that is used to further ascertain the high redshift nature of the
object. The work of Steidel et al. (\cite{steidel1},
\cite{steidel2},\cite{steidel3}) imaging fields around high redshift
quasars yielded many candidates, several of which have been 
spectroscopically confirmed to be galaxies at $z\approx 3$ 
The advantage over previous unsuccessful narrowband Lyman-$\alpha$ surveys 
(e.g. Giavalisco et al. \cite{giavalisco}, Mart\'\i nez-Gonz\'alez et al.
\cite{kike}) is that a wide range in redshift can be surveyed. Besides, the
break in the spectrum can be easily discerned with broad band imaging, allowing 
the detection of normal high-$z$ galaxies, with a typical magnitude at a
redshift of $z\sim 3$--$4$ of $R\sim 25^m$ (Djorgovski \cite{djorgovski}). 
Furthermore, the Lyman-$\alpha$ emission line can be weak or even non-existent 
as it appeared in many of the galaxies detected by Steidel et al. \cite{steidel4}.

We found the field around quasar PC1643+4631A to be a good candidate 
for this technique for a couple of reasons: This $z=3.79$
QSO has a Damped Lyman-$\alpha$ System (DLS) 
at $z=3.14$ (Schneider et al. \cite{schneider})
whose Lyman break would be perfectly positioned between Johnson 
$U$ and $V$ filters. Besides, a possible very massive cluster might
exist between this quasar and a second one (PC1643+4631B) that lies 
just 200 arcsec away and has roughly the same redshift 
($z=3.831$). Jones et al. \cite{jones} detected a significant 
decrement in the CMB in a region
between these two QSOs that could be related with a Sunyaev-Zeldovich
effect arising from a M$\sim 10^{15}$M$_\odot$ intervening cluster. 
However, Saunders et al. \cite{saunders} imaged the region in $R$,$J$ and
$K$ bands, looking for ellipticals that should abound in such a cluster,
and found no evidence. The technique used in this paper is the best one 
to search for galaxies in the redshift range $z\sim 3$ -- $4$, thereby
contributing to check this result: If such a massive cluster were to 
exist, a survey around PC1643+4631A should give a net excess 
of candidates towards the direction of the cluster.

\begin{table}
  \caption[]{Imaging around PC1643+4631A}
  \label{Img}
  \begin{tabular}{c|c|cc}\hline
  Band & Exposure Time (sec.) & Seeing & Lim.~Mag.\\
  \hline\hline
  U & 18000 (NOT) + 2400 (WHT) & 1.2$^{\prime\prime}$ & 26.0 \cr
  V & 		    3600 (WHT) & 1.0$^{\prime\prime}$ & 26.0 \cr
  R & 7200 (NOT) + 1800 (WHT)  & 1.5$^{\prime\prime}$ & 26.0 \cr
  \hline\hline
  \end{tabular}
 \end{table}

\section{Observations}

The observations were obtained in May 1996 using BroCam2 at the 2.5 m Nordic 
Optical Telescope (NOT) and in May 1997 using the William Herschel Telescope
(WHT) 4.2 m prime focus camera, both at the Roque de los Muchachos 
Observatory, La Palma, Spain.
Both cameras used a thinned $2048\times 2048$ Loral chip which is
remarkable for its very high quantum efficiency even in the blue part 
of the spectrum, making it especially useful for $U$ band imaging.
The image scale is $0.11^{\prime\prime}$ and $0.265^{\prime\prime}$ 
per pixel (NOT and WHT respectively). 
We also used a 30 minute exposure of the quasar in the $R$ band 
(WHT+TEK CCD detector) taken by Saunders....., retrieved
from the ING archive, which greatly helped us in achieving a homogenous
limiting magnitude in all three bands.
Deep imaging was done through 
Johnson $U$, $V$ and $R$ filters and reduced using the standard techniques of
bias subtraction and flat fielding using IRAF tasks. The long integration times
required were separated into shorter --- usually 1200 second --- exposures,
slightly nodding the telescope between them in order to be able to eliminate
cosmic rays and bad pixels.
These frames were eventually registered and combined into a final
image using a clipping algorithm. The photometric calibration was 
achieved with images of standard stars (Landolt \cite{landolt})
taken throughout the observing nights. 

In order to detect candidates of galaxies at high redshift, a very deep
exposure must be achieved in each filter. The deepest band should be $U$
because it represents the spectral range shortward of the Lyman break
for $z\approx 3$ objects.

 \begin{table}
   \caption{Lyman Break Candidates within 1.5 arcmin}
\bigskip
   \label{candidates}
   \begin{tabular}{r|rr|crr}\hline
No. & $\Delta\alpha(^{\prime\prime})$ & $\Delta\delta(^{\prime\prime}) $ 
& $V$ & $U-V$ & $V-R$\\
\hline\hline
  1 &  -0.3 & -6.6 & 24.09 &   0.97 &  0.78\cr
  2 &  -5.6 &  7.4 & 25.00 &$>$1.00 & -0.16\cr
  3 & -22.2 & 10.8 & 24.64 &$>$1.36 & -0.11\cr
  4 &   8.4 &-29.9 & 24.57 &   0.77 & -0.24\cr
  5 & -27.0 & 29.0 & 25.22 &$>$0.78 & -0.27\cr
  6 & -21.3 &-34.2 & 24.57 &$>$1.43 &  0.12\cr
  7 &   1.4 & 42.6 & 25.36 &$>$0.64 &  0.43\cr
  8 &  63.2 & -0.7 & 24.96 &   0.70 &  0.14\cr
  9 &   9.9 &-43.5 & 25.30 &$>$0.70 &  0.14\cr
 10 & -64.9 & -3.2 & 24.09 &   0.90 & -0.01\cr
 11 &  59.1 &-29.4 & 24.52 &$>$1.48 & -0.18\cr
 12 &  31.4 &-46.9 & 24.67 &$>$1.33 &  0.46\cr
 13 &   8.4 &-54.5 & 24.79 &$>$1.21 & -0.00\cr
 14 &  28.2 &-57.0 & 24.30 &   0.77 & -0.21\cr
 15 &  58.6 & 49.0 & 24.98 &$>$1.02 &  0.09\cr
 16 & -66.1 &-44.5 & 25.24 &$>$0.76 &  0.01\cr
 17 &  79.3 & 33.2 & 25.23 &$>$0.77 &  0.81\cr
 18 & -23.7 & 71.3 & 24.72 &$>$1.28 & -0.20\cr
 19 & -12.4 & 74.0 & 24.24 &   0.92 &  0.22\cr
 20 & -82.7 &-55.3 & 25.43 &$>$0.57 & -0.17\cr
 21 & -18.9 & 82.5 & 24.80 &   0.45 &  0.67\cr
\hline\hline
  \end{tabular}
 \end{table}

\section{Analysis}

The search of objects in the final frames was performed using SExtractor
(Bertin E. \cite{bert}), carefully tweaking the parameters to avoid such 
effects as excessive deblending or noise--dominated detections. 
The threshold was set at
16 connected pixels (which is slightly smaller than the number of pixels
inside the seeing disk) with 0.75$\sigma$ lower limit per pixel for a 
global 3$\sigma$ detection per object. We performed aperture photometry on
the objects using several aperture sizes to make sure we were collecting
all the flux. It also gave us an idea of the uncertainty to expect with faint
objects. Finally, we chose a diameter of 12 pixels (3.2 arcsec) to get the
photometry and checked
each candidate visually to make sure there was no problem with deblending,
hot pixels or other spurious detections. This issue was further enforced by 
considering only candidates that appeared {\sl both} in the $V$ and $R$ filter
images. The size of the aperture was chosen taking into account the seeing of the
images (1--1.5 arcsec) and the fact that high--$z$ candidates should be
pointlike objects.

\begin{figure}
\resizebox{\hsize}{!}{\includegraphics{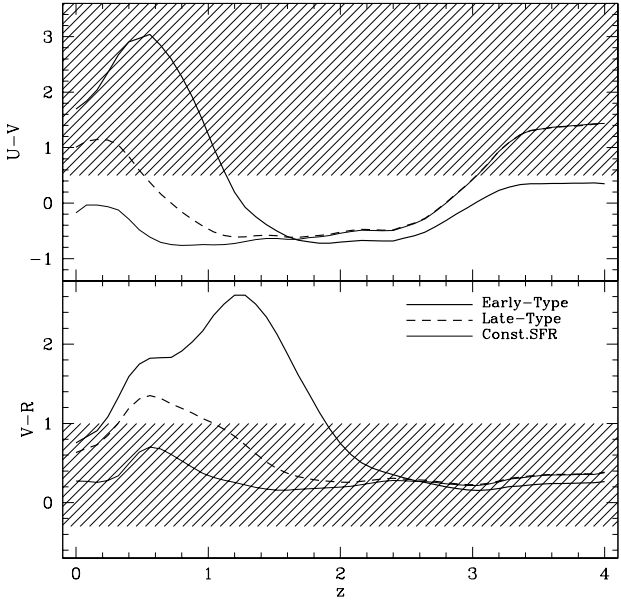}}
\caption{
	Predictions for three different morphologies from the spectral 
	evolution models of Bruzual \& Charlot (\cite{bruzual}). The shaded
	areas represent the color constraints when choosing candidates for
	high-redshift galaxies.
}
   \label{model}
\end{figure}

The photometry output from SExtractor
was further checked using several tasks from IRAF from which we can
extract a photometric uncertainty of 0.3--0.4 magnitudes for objects
with $V>24^m$. The candidates for high redshift galaxies were obtained
using a color and magnitude constraint. The first and most important
one is $U-V>0.5$ since it reveals the existence of a break in the spectrum
at around 3650\AA\  which corresponds to the Lyman continuum break at 
$z\sim 3$. 
Unfortunately, there is also a very strong break at 4000\AA\  
in the rest frame, especially prominent in old galaxies. Hence, low-$z$
early--types can sneak in our list. Furthermore, we could also find 
galaxies with a strong emission line falling in the $V$ band and 
faking such a break. The way out of this degeneracy is twofold: 
we check an extra color towards the red side of the spectrum, 
namely $V-R$, which can tell low redshift ellipticals or galaxies 
with strong emission lines from high redshift ones. We took 
$-0.3\leq V-R \leq 1.0$. Figure~\ref{model} shows the color prediction 
for three different morphologies from the models of 
Bruzual \& Charlot \cite{bruzual}.
Notice how the degeneracy between high redshift late--type galaxies and
low redshift ellipticals in the $U-V$ graph is removed when considering
$V-R$. 

An early--type galaxy has a steeper spectrum towards $R$, thereby
higher values of $V-R$, whereas the lower bound in this color is chosen
to avoid galaxies with strong emission lines which might fall in the 
$V$ band. Besides the color constraints, we should also impose a limit on the 
apparent magnitude. Even though the selection suffers from Malmquist's
bias, we do not expect to find galaxies with an intrinsic luminosity
greater than a few $L_\star$, which implies rejecting candidates whose
apparent magnitude is brighter than $V\sim 24^m$. Part of the degeneracy 
may still remain because faint galaxies could still be
identified with ellipticals at low redshift with an absolute luminosity
around $L<0.04L_\star$. However, the constraint in $V-R$ makes this choice
less likely over a late--type high redshift galaxy.

\begin{figure}
\resizebox{\hsize}{!}{\includegraphics{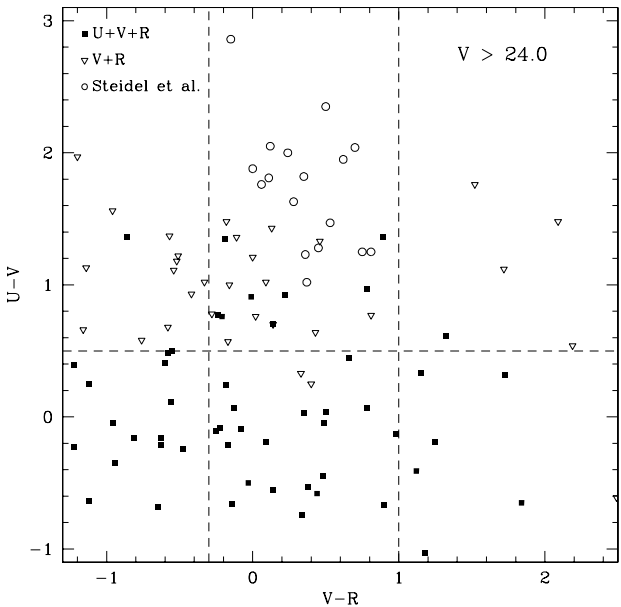}}
\caption{
	Color--Color diagram of faint ($V>24^m$) objects detected around
        PC1643+4631A. The dashed lines engulf the region inside which
        proper candidates must lie. These objects use different symbols
        depending on whether there has been a matching detection in the
        $R$ or $U$ band (as well as the $V$ band which is used as
        reference). The confirmed $z\approx 3$ galaxies listed in Steidel 
        et al. (1996) are also shown.}
   \label{objects}
\end{figure}

\begin{figure}
\resizebox{\hsize}{!}{\includegraphics{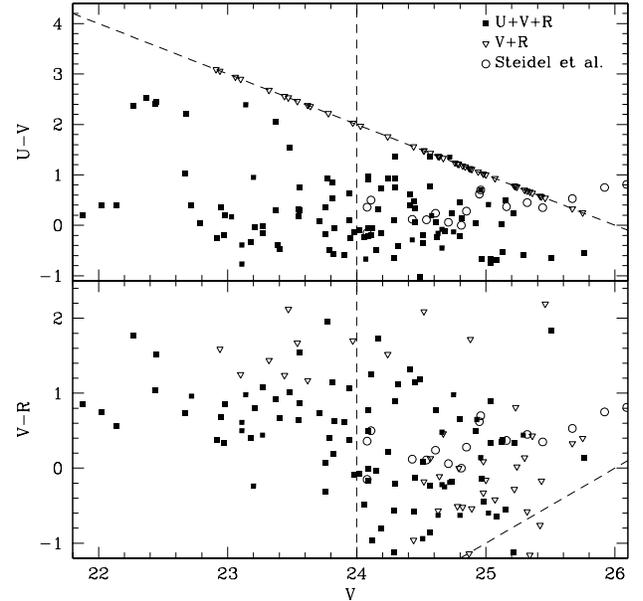}}
\caption{
	Color--Magnitude diagrams for objects around PC1643+4631A
	detected both in $V$ and $R$. 
	The dashed lines represent the limiting magnitude. Steidel et al.'s
	confirmed candidates are also shown.}
   \label{allobjs}
\end{figure}

Table 2 shows the list of candidates which appeared in our images 
imposing these constraints. All the objects with detections both in 
$V$ and $R$ are shown in Figure~\ref{allobjs}, whereas 
Figure~\ref{objects} plots a subsample of these objects with $V>24.0^m$
and shows the constraining color--color region where the candidates should
lie. 
Figure~\ref{objects}  also shows the confirmed candidates from Steidel et al.
(\cite{steidel4}), who used the same technique although with a 
different set of filters, specifically designed for maximum efficiency
at these redshifts. 

Finally, we considered the object that lies very close to the line of sight
of the QSO. The $V$ and $R$ images show it as an elongation of the quasar
towards the SW, but in the $U$ band it stands out clearly as the QSO
fades out of view. In order to perform the photometry in $V$ 
and $R$, we subtracted the PSF of a nearby point-like object scaled to the
peak of the QSO. This is just a rough estimate but it allows us to ascertain
the nature of this object. We got $V = 22.7$; $U-V = 0.3$ and $V-R = 1.6$.
It is rather bright in the $U$ band, which discards the possibility of 
it being the one responsible for the DLS in the QSO's spectrum at $z=3.14$. 
The models fit these colors with a late--type galaxy at $z\sim 0.5$, which
--- for a $\Omega_0=1$, $h_0=0.5$ cosmology --- gives an absolute luminosity
of $L\sim 0.2L_\star$. This galaxy might possibly leave its imprint on the
spectrum of the QSO.

\section{Hu \& Ridgway's red objects revisited}

Among the many objects detected around PC1643+4631A, it is worth 
mentioning a group of them that are very red in $U-V$ but too bright
to be high redshift galaxies. Many of these objects were already
detected by Hu \& Ridgway (\cite{hu}), using a selection criteria
in the red and NIR part of the spectrum. They looked for objects
which had a very red $I-K$ color, hence aiming at the detection of
early--type galaxies and EROS. 
The 19 objects listed in their work are shown in table 3 along
with our photometry in $U$,$V$ and $R$. Combining both surveys we
have photometry in 8 standard filters ($U,B,V,R,I,J,H,K^\prime$) 
which can be used to determine the nature of some of these objects.
Most of them have very red $U-V$ and $R-K^\prime$ colors as well as
relatively bright $V$ magnitudes, all evidence pointing at low redshift
elliptical galaxies. The high density of these objects may plausibly
imply the existence of a cluster. Furthermore, radio--loud QSOs --- such
as PC1643+4631A --- have a higher probability of being magnified by
foreground matter (i.e. an intervening cluster) because of the double
magnification bias (Ben\'\i tez \& Mart\'\i nez-Gonz\'alez \cite{beni}, 
Schneider et al. \cite{sch}). The last entry in table 3 gives
an estimated redshift using the models of Bruzual \& Charlot 
(\cite{bruzual}). The square difference between the actual and the
predicted models is computed and the $z$ for which this difference
reaches a minimum gives the estimated redshift. The letter in front
labels the type of galaxy model that fits best: E and L for early
and late type, respectively.

Objects \# 10 and 14 cannot be catalogued this way because they escape 
detection down to very faint magnitudes ($U>26^m$ and $V>26^m$). They
are very good candidates for EROS and that prompted the observations of
object \# 10 doing deep IR imaging and spectroscopy 
(Graham \& Dey \cite{graham}), finding --- as expected --- that 
it could not be an elliptical galaxy but rather  an interacting system.
A possible H$\alpha$ emission at $z\approx 1.44$ was detected, revealing
quite a strong star formation rate.

\begin{table}
   \caption{Red Objects from Hu \& Ridgway 1994. The quasar is object \# 7.}
 \bigskip
   \label{hrobj}
   \begin{tabular}{c|c|cccc|c}\hline
No. & $V$ & $U-B$ & $B-V$ & $V-R$ & $R-K^\prime$ & $z_{\rm FIT}$\\
\hline\hline
  1	& 20.8    & 1.4		& 1.3	& 2.4	& 2.7 &  E/0.5\cr
  2	& 19.3    & 1.3		& 1.2	& 1.2	& 2.2 &  E/0.4\cr
  3	& 20.2    & -0.6	& 1.9	& 1.0	& 3.1 &  ---\cr
  4	& 19.8    & -1.1	& 1.7	& 0.7	& 2.6 &  ---\cr
  5	& 20.6    & 1.4		& 1.3	& 1.5	& 2.5 &  E/0.4\cr
  6	& 21.4    & 0.0		& 1.5	& 1.5	& 2.7 &  L/0.4\cr
  7     & 20.2	  &$>$4.6	& 1.2	& 0.7   & 2.2 &  ---\cr
  8	& 21.8    & $>$2.7	& 1.0	& 0.9	& 2.9 &  E/0.4\cr
  9	& 23.8    & -0.6	& 1.6	& 2.0	& 3.6 &  E/1.0\cr
 10	& $>$26.0 & $>$-1.0	&$<$0.5	& ---	& $>$6.6 &  ---\cr
 11	& 23.0    & $>$0.7	& 1.8	& 2.0	& 2.6 &  E/0.4\cr
 12     & 24.1    & -0.8	& 0.7	& 1.3	& 4.4 &  ---\cr
 13	& 24.0    & $>$0.3	& 1.2	& 1.9	& 3.6 &  E/0.4\cr
 14	& $>$26.0 & ---		& ---	& ---	& $>$6.3 &  ---\cr
 15	& 21.4    & -1.0	& 1.3	& 0.1	& 2.5 &  ---\cr
 16	& 23.0    & -1.2	& 0.9	& 0.4	& 3.7 &  ---\cr
 17	& 23.4    & $>$0.2	& 1.9 	& 1.5	& 3.0 &  E/0.3\cr
 18	& 22.7    & 0.0 	& 1.7	& 0.3	& 3.5 &  ---\cr
 19	& 22.4    & $>$1.9	& 1.2	& 0.9	& 2.5 &  E/0.4\cr
\hline\hline
   \end{tabular}
 \end{table}

\section{Conclusions}
The search for the Lyman break in high redshift galaxies is so far
the most efficient way of selecting candidates for follow-up 
spectroscopy.
The field around PC1643+4631A ($z=3.79$) is a very attractive
target for this method since it has a DLS at $z=3.14$ which sets the
Lyman break between Johnson $U$ and $V$ frames. We performed deep imaging
in $U$,$V$ and $R$ and obtained a list of candidates for late--type
high redshift galaxies. The claim of Jones et al. (\cite{jones})
that a massive cluster might lie close to this QSO 
--- based on a Sunyaev-Zeldovich detection --- cannot be confirmed as
the listed candidates are distributed homogeneously over the frame.
Our result corroborates the work of Saunders et al. (\cite{saunders}), who
did not detect any overabundance of objects in NIR images (aiming
at the detection of early--type galaxies).

\begin{figure}
\resizebox{\hsize}{!}{\includegraphics{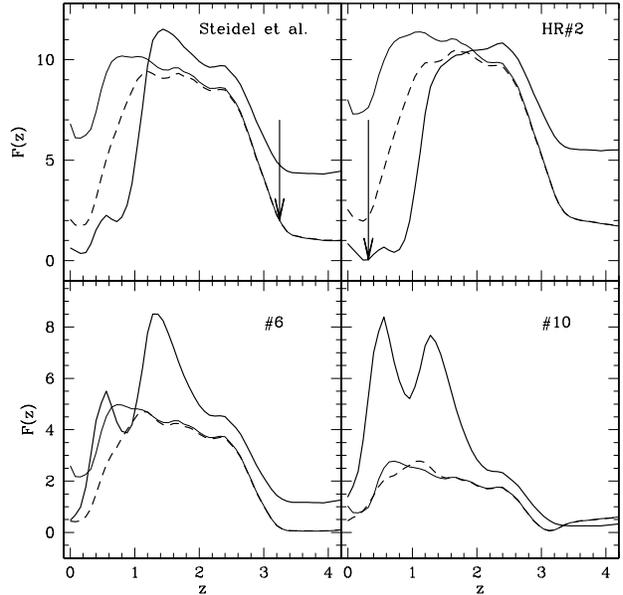}}
\caption{
	Color--color fit to spectra from the models of Bruzual \&
	Charlot (1997) for (clockwise from upper left) one of Steidel
	et al's (1996) confirmed high-$z$ galaxies, one of Hu \& Ridgway's
	(1994) red object, and a couple of our candidates. 
	F(z) represents the sum of the squares of the
	differences in $U-V$ and $V-R$ between the actual colors and
	the ones from the models. The thick,dashed and thin lines 
	represent early--type, late--type and constant SFR, respectively.
	The arrow in Steidel et al.'s box show the actual redshift of the
	object, the one in Hu \& Ridgway's box show the expected low redshift
	value which is also considered against high-$z$ from the bright
	nature ($V=19.3$) of the object.}
   \label{fits}
\end{figure}

Besides, this work has completed the data from Hu \& Ridgway 
(\cite{hu}) for a sample of 18 objects around the quasar which have
a very red $I-K$ color. The addition of $U$, $V$ and $R$ photometry
to these objects allows a better identification. We found many of them
to match the colors for early--type normal galaxies at $z\sim 0.4$.
The extremely red objects \# 10 and 14 in the list from Hu \& Ridgway's
work are fainter than the limiting magnitude of our search 
($U,V,R>26.0$) which means these two objects must
have a very steep rise towards the NIR part of the spectrum.

\begin{acknowledgements}
The WHT and NOT are operated on the island of La Palma by the Royal Greenwich
Observatory and the Nordic Optical Telescope Scientific Association, 
respectively, in the Spanish Observatorio del Roque de los Muchachos of the 
Instituto de Astrof\'\i sica de Canarias.
We would like to thank the Isaac Newton Group of Telescopes 
for such a friendly interface to retrieve images from its archive.
I.F. and N.B. acknowledge a Ph.D. scholarship from the 
`Gobierno de Cantabria', and
the Spanish MEC, respectively. I.F.,N.B. and E.M.-G. acknowledge financial 
support from the Spanish DGES under contract PB95-0041.
\end{acknowledgements}


\end{document}